\documentclass[aps,pra,twocolumn]{revtex4-2}

\usepackage[utf8]{inputenc}
\usepackage{amsmath}
\usepackage{stmaryrd}
\usepackage{color}
\usepackage{mathrsfs}
\usepackage{yfonts}
\usepackage{rotating}
\usepackage{graphicx}
\usepackage{amsfonts}
\usepackage{mathrsfs}
\usepackage{amsmath}
\usepackage[american]{babel}
\usepackage{wasysym}
\usepackage{slashed}
\usepackage{pgfplots}
\usepackage{braket}
\usepackage{graphicx}
\usepackage[caption=false]{subfig}
\usepackage{dsfont}
\usepackage{amsthm}
\usepackage{youngtab}
\usepackage{gensymb}

\usepackage{natbib}
\PassOptionsToPackage{breaklinks}{hyperref}
\usepackage[colorlinks,allcolors=black]{hyperref}

\theoremstyle{definition}

\usepackage{cleveref}
\crefname{proposition}{Proposition}{Propositions}
\crefname{theorem}{Theorem}{Theorems}
\crefname{definition}{Definition}{Definitions}
\crefname{lemma}{Lemma}{Lemmas}
\crefname{figure}{Figure}{Figures}
\crefname{corollary}{Corollary}{Corollary}
\crefname{conjecture}{Conjecture}{Conjectures}
\crefname{section}{Section}{Sections}
\crefname{appendix}{Appendix}{Appendixes}
\crefname{observation}{Observation}{Observation}
\crefname{remark}{Remark}{Remark}
\crefname{example}{Example}{Examples}
\crefname{equation}{Eq.}{Eqs.}
\crefname{table}{Table}{Tables}

\usepackage{tikz}

\usepackage{amssymb}
   \makeatletter
     \renewcommand\@make@capt@title[2]{%
      \@ifx@empty\float@link{\@firstofone}{\expandafter\href\expandafter{\float@link}}%
       {\textbf{#1}}\@caption@fignum@sep#2\quad}%
     \makeatother
\makeatletter
\renewcommand{\fnum@figure}{\textbf{Figure~\thefigure}}

\usepackage{mathtools}

\newcommand{\de}[0]{\delta}

\newcommand{\la}[0]{\lambda}

\newcommand{\si}[0]{\sigma}

\newcommand{\bs}[1]{\textbf{#1}}
\newcommand{\ea}[1]{\begin{align}#1\end{align}}
\newcommand{\eq}[1]{\begin{equation}#1\end{equation}}

\newcommand{\ma}[1]{\mathcal{#1}}

\usepackage{cancel}

\begin{document}


\title{Multipartite Entanglement from Consecutive Scatterings}
\author{Gon\c{c}alo M. Quinta}
\email{goncalo.quinta@tecnico.ulisboa.pt}
\affiliation{Instituto de Telecomunica\c{c}\~{o}es, Lisboa, Portugal}
\author{Rui André}
\email{rui.andre@tecnico.ulisboa.pt}
\affiliation{Instituto de Telecomunica\c{c}\~{o}es, Lisboa, Portugal}


\begin{abstract}

We study how the successive scattering of spin 1/2 particles with a central spin 1/2 target particle can generate entanglement between the helicity degrees of freedom of all scattered particles, effectively producing a multipartite entangled state. We show that the bipartite entanglement between each pair of scattered particles, as quantified by the concurrence, is largest for reflected particles and decreases with the number of scatterings. We study the entanglement generation as a function of the scattered particles momenta, angular distribution and mass ratios, and show that there is always a combination of optimal helicities and momentum which generate the largest amount of bipartite entanglement.

\end{abstract}

\maketitle


\section{Introduction}

The development rate of quantum technologies is inhrently dependent on the ability to generate and manipulate entanglement in quantum states. Despite the many decades necessary to manifest entanglement in the laboratory, it is now considered a physical resource able to be, among other things, manipulated \cite{PhysRevLett.74.2619, PhysRevLett.76.722, PhysRevA.54.3824, RevModPhys.73.565} and distributed \cite{doi:10.1080/09500340008232183, 10.1063/1.1712500, PhysRevA.75.022327}. The vast majority of quantum information applications, such as dense coding, quantum teleportation, quantum key distribution \cite{Traub2009}, error correcting codes \cite{PhysRevLett.77.198} or quantum computation \cite{nielsen_chuang_2010, JozsaLinden}, are in fact, in one way or another, a particular way to exploit the properties of quantum entanglement. 

To generate entanglement, one needs to make at least two systems interact in such a way that their quantum degrees of interest are not measured after the interaction. The most natural setup in which this occurs is during particle colisions. In fact, any quantum mechanical interaction is most fundamentally described by Quantum Field Theory (QFT), so it is expected that the generation of entanglement is rooted in the interactions of elementary particles, scatterings being the most common ones. Research on the role of entanglement generation has gained particular traction in recent years in several fields of high-energy physics, such as neutrino oscillations \cite{Blasone_2009, PhysRevD.82.093003, BLASONE2013320, Blasone_2014, quinta2022predicting}, Quantum Electrodynamics (QED) \cite{SciPostPhys.3.5.036, PhysRevD.107.116007, PhysRevD.100.105018} and Quantum Chromodynamics \cite{Afik2022quantuminformation}. However, particle scatterings have gained a particular emphasis due to their conceptual simplicity and richness. The typical physical setup consists in considering the helicity quantum degrees of freedom of two initially separated particles which become entangled after the collision. Using Feynman diagrams, it is straightforward to investigate all manners of entanglement generation for different types of interaction. The generation of entanglement specifically through particle scatterings has been studied in QED in \cite{SciPostPhys.3.5.036}, where maximal entanglement between helicity degrees of freedom was shown not only to be present in most possible scatterings but also possibly related to gauge symmetry of QED. A more in-depth treatment of entanglement generation in QED scatterings was also performed in \cite{PhysRevD.107.116007}. In a context of more than two particles, it was shown recently \cite{PhysRevD.100.105018}, that the cross sectional information between two-particle colisions can be encoded in a third spectator particle, as long as it is initially entangled with one of the scattered particles. Overall, 2-particle scattering scenarios have been reasonably explored but scatterings of multiple particles have not been addressed to date.

Particle scaterings typically happen in a small time window, which makes it very unlikely that more than two particles actually collide at the same time in such a way that multiple particles need to be considered in a single Feynamn diagram. Nevertheless, multiple consecutive 2-particle scatterings can occur before any of the final states is colapsed. This leads to a picture where particles in entangled states coming from past scatterings can collide with new separable particles, resulting in increasingly larger multiparticle entangled states.

In this work, we explore this simple scenario and analyse its usefullness as a method to create multipartite entangled states. In particular, we consider a spin 1/2 ``target'' particle initially at rest and subject it to consecutive collisions with lighter ``projectile'' particles coming from an external source. Although we do not develop on the experimental side, this works provides a new conceptual method that indeed generates multipartite states from collisions and shows what entanglement properties one can expected from such states.


\section{Entanglement in Particle Scatterings}

We begin by specifying the nomenclature used throughout this work and show how entanglement is naturally present in the quantum degrees of freedom involved in particle scatterings, specifically for the fermionic case. Consider $n$ particles with masses $m_{i}$ at a certain time instant $t=t_i$, labelled by their 3-vector momenta $\bs{p}_{i}$ and basis indexes $h_{i}$, where henceforth we will use bold notation for 3-vectors. The $h_{i}$ can be related to spin or helicity, for example. For simplicity, we shall restrict this work to the case of equal number of initial and final particles as well as the conservation of the number of each particle species present in the reaction. We assume that at $t_i$ they are sufficiently far away such that together they are in an eigenstate of the free Hamiltonian, written in the form
\eq{
\ket{\psi(t_i)} = \ket{\bs{p}_1, h_1; \ldots; \bs{p}_n, h_n}\,.
}
After a certain period of time they interact and scatter, and at a further enough time instant we consider them again to be in an eigenstate of the free theory. This evolution is encoded in the scattering matrix, given by
\eq{\label{evolutionU}
\hat{U} = T\left\{ \exp\left( -i \int^{+\infty}_{-\infty} \hat{V}(t') dt' \right) \right\}\,,
}
where $T$ is the time-ordering operator and $\hat{V}(t')$ is the potential part of the Hamiltonian in the interaction picture and a hat notation will be adopted for operators. The scattered state is thus of the form
\eq{\label{finalstate}
\ket{\psi(t_f)} = \hat{U}\ket{\psi(t_i)} = (1+i\hat{\ma{T}})\ket{\psi(t_i)} \,,
}
where we used the standard decomposition $\hat{U} = 1 + i \hat{\ma{T}}$. We will use the standard nomenclature
\ea{
\bra{\bs{q}_1, r_1; \ldots; \bs{q}_n, r_n} i\hat{\ma{T}} \ket{\bs{p}_1, h_1; \ldots; \bs{p}_n, h_n} =  \nonumber \\
& \hspace{-60mm} (2\pi)^4 \de^{(4)}\left(\sum q_i - \sum p_i\right) \times \nonumber \\
& \hspace{-55mm} \times i \ma{M}(\bs{q}_1, r_1; \ldots; \bs{q}_n, r_n | \bs{p}_1, h_1; \ldots; \bs{p}_n, h_n)
}
for the non-trivial piece of the scattering amplitude from the initial state $\ket{\psi(t_i)}$ to a n-particle state with momenta $\bs{q}_{i}$ and basis index $r_{i}$, where the Dirac delta ensures momentum conservation. The crucial point of Eq.~(\ref{finalstate}) is that, so long as the final dynamical quantities are not measured, the final state will be in a superposition of all possible outgoing states. This can be seen by applying the $n$-particle identity operator $\hat{I}_n$ from the left in Eq.~(\ref{finalstate}), where
\ea{
\hat{I}_n & = \int \frac{d^3 \bs{q}_1}{(2\pi)^3 2E_{\bs{q}_1}} \ldots \frac{d^3 \bs{q}_n}{(2\pi)^3 2E_{\bs{q}_n}} \nonumber \\
& \sum_{r_1, \ldots, r_n} \ket{\bs{q}_1, r_1; \ldots; \bs{q}_n, r_n}\bra{\bs{q}_1, r_1; \ldots; \bs{q}_n, r_n}\,,
}
where $d^3 \bs{q} \equiv dq_x dq_y dq_z$ is the spatial volume element and $E_{\bs{p}_i} = \sqrt{|\bs{p}_{i}|^2 + m_{i}^2}$ are the particles' energies. This leads to the following form for the final state:
\ea{\label{genFinalState}
\ket{\psi(t_f)} & = \ket{\psi(t_i)} \nonumber \\
& \hspace{-9mm} + i (2\pi)^4 \int \frac{d^3 \bs{q}_1}{(2\pi)^3 2E_{\bs{q}_1}} \ldots \frac{d^3 \bs{q}_n}{(2\pi)^3 2E_{\bs{q}_n}} \de^{(4)}\left(\sum q_i - \sum p_i\right) \nonumber \\
& \times \sum_{r_1, \ldots, r_n}  \ma{M}(\bs{q}_1, r_1; \ldots; \bs{q}_n, r_n | \bs{p}_1, h_1; \ldots; \bs{p}_n, h_n) \nonumber \\
& \times \ket{\bs{q}_1, r_1; \ldots; \bs{q}_n, r_n}
}
which clearly shows that the final state is a superposition of all possible $\ket{\bs{q}_1, r_1; \ldots; \bs{q}_n, r_n}$. Such a superposition will in general be impossible to decompose as a tensor product, indicating that the final state will be entangled in general.

Finally, it is important to note that the initial state $\ket{\psi(t_i)}$ can also be entangled from the very beginning, even if the particles are very far away from each other. Indeed, such a group of particles might have originated from a different scattering and as a consequence became entangled. This is also fully consistent with the free theory eigenstate assumption. To see this, consider the general free theory Hamiltonian written in second quantized form
\eq{
\hat{H}_{0} = \int \frac{d^3 \bs{p}}{(2\pi)^3} \sum_{h} E_{\bs{p}} \hat{a}^{h \dagger}_{\bs{p}}\hat{a}^{h}_{\bs{p}}
}
where $\hat{a}^{h}_{\bs{p}}$ is the annihilation operator, i.e. the anticommuting operator which annihilates the free theory vacuum $\ket{0}$ and obeys the relation
\eq{\label{commrel}
\{\hat{a}^{h}_{\bs{p}},\hat{a}^{h' \dagger}_{\bs{p}'}\} = \delta_{h h'} \delta(\bs{p}-\bs{p}')\,.
}
The initial state can then be written as
\eq{
\ket{\bs{p}_1, h_1; \ldots; \bs{p}_n, h_n} = (2\pi)^n \hat{a}^{h_1 \dagger}_{\bs{p}_1} \ldots \hat{a}^{h_n \dagger}_{\bs{p}_n} \ket{0}\,.
}
Focusing on the helicity quantum degrees of freedom, which are what we shall be concerned with in this work, if we consider a general linear combination of the form
\eq{
\ket{\tilde{\psi}} = \sum_{h_1 \ldots h_n} c_{h_1 \ldots h_n} \ket{\bs{p}_1, h_1; \ldots; \bs{p}_n, h_n}\,,
}
one can straightforwardly conclude from Eq.~({\ref{commrel}}) that
\eq{
\hat{H}_{0} \ket{\tilde{\psi}} = \left(\sum_i E_{\bs{p}_i}\right)\ket{\tilde{\psi}}
}
and so one proves that $\ket{\tilde{\psi}}$ is also an eigenstate of the free theory. Since $\ket{\tilde{\psi}}$ represents a general form of an entangled state, one thus concludes that entangled states are also eigenstates of the free theory.


\section{A single scattering}

We shall focus first on the entaglement generated from a single scattering, taking the limit where both projectile and target particles are point-like with no internal structure. 
The setup consists of a target particle of rest-mass $M$ with 4-momentum $p_1 = (M,0,0,0)$, and a projectile particle of rest-mass $m$ aligned with the $z$-axis with 4-momentum $p_2 = (\sqrt{m^2+p^2}, 0,0,p)$ fired at the target particle. 
Ideally, the target particle is heavy enough (compared to the projectile one) for the recoil to be considered negligible, in order to experimentally facilite multiple collisions with the same heavy particle. From the theoretical perspective of this work, we consider that each projecile particle always hits the target particle. The direction of the outgoing projectile particles is determined by the angles $\theta$ and $\phi$, corresponding to the azimuthal and polar angles respectively, where from polar symmetry we can fix $\phi=0$ in the first collision. The quantum degrees of freedom whose entanglement we would like to study are the helicities of both particles, which start out in a separable state. Denoting the initial and final helicities of the projectile and target particles as $h_1, h_2$ and $h_3, h_4$, respectively, one can apply the unitary evolution operator $\hat{U}_{1,2}$ defined as $\hat{U}$ from Eq.~(\ref{evolutionU}) acting on two particles and contract the result with $\bra{\bs{p}_3,h_3;\bs{p}_4,h_4}$. This leads to the transition probability for the case where the scattering occurs, given by (see Sec.~A of the Appendix)
\ea{
\bra{\bs{p}_3,h_3;\bs{p}_4,h_4}\hat{U}_{1,2} \ket{\bs{p}_1, h_1; \bs{p}_2, h_2} = \nonumber \\
& \hspace{-50mm}  \ma{N}_2 \, \ma{M}(\bs{p}_1+\bs{p}_2-\bs{p}_4, h_3; \bs{p}_4, h_4 | \bs{p}_1, h_1; \bs{p}_2, h_2) 
}
where 3-momentum conservation is evident and the normalization factor
\eq{
\ma{N}_2  = i (2\pi)^4 \de\left(E_{\bs{p}_3}+E_{\bs{p}_4}-E_{\bs{p}_1}-E_{\bs{p}_2}\right)
}
ensures conservation of energy. Focusing on the helicity degrees of freedom, one finds that the final state is of the form
\eq{
\ket{\psi_f} = \sum_{h_3,h_4} \ma{M}(h_3, h_4 | h_1, h_2) \ket{h_3,h_4}
}
where we omit the momentum notation and superfluous constant factors. Momentum and energy conservation are always implicit, despite the omition of momentum vector variables. The state $\ket{\psi_f}$ is a 2-qubit state, where we consider the index 0 and 1 to be associated to the helicities $+1$ and $-1$, respectively. The entanglement between the helicty (qubit) degrees of freedom can be quantified by the concurrence \cite{PhysRevA.61.052306}, denoted here by $\ma{C}$. This quantity can be expressed as
\eq{\label{eqconcurrence}
\ma{C}(\rho) = \textrm{max}(0,\la_1-\la_2-\la_3-\la_4)
}
where $\la_1, \ldots, \la_4$ are the eigenvalues, in decreasing order, of the Hermitian matrix
\eq{
R = \sqrt{\sqrt{\rho} \tilde{\rho} \sqrt{\rho}}
}
with
\eq{
\tilde{\rho} = (\si_y \otimes \si_y) \rho^{*} (\si_y \otimes \si_y)\,,
}
where $\si_y$ are the Pauli matrices in the $y$ direction and $\rho$ is the density matrix of the system.
For simplicity, we will denote the concurrence between particles $\text{A}$ and $\text{B}$ as $\mathcal{C}[\text{A},\text{B}]$. Although any superposition of helicities can be considered as an initial state, it is more convenient to start with a separable state in order to to investigate how much entanglement was generated. 
The only helicity state that can yield maximal entanglement, i.e. $\mathcal C = 1$, is that of opposite helicities, such as $\ket{01}$.
Consequently, it will be the case of focus in this section.
Additionally, we will focus on the entanglement being generated on the vicinity of $\theta=\pi$, as it has already been seen~\cite{PhysRevD.107.116007} that this is the most prolific case for entanglement generation in $t$-channel QED scatterings.

\begin{figure}[h]
\centering
\includegraphics[width=\columnwidth]{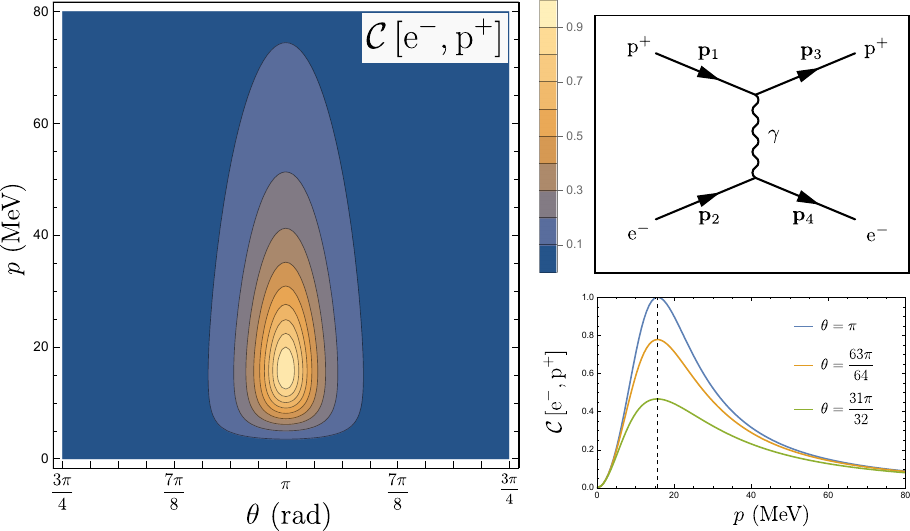}
\caption{Top right: The Feynman diagram of the interaction. Left half and bottom right: Concurrence generated from scattering an electron off a proton at rest in the laboratory frame. Choosing as initial state a polarized state where both particles have opposing helicities yields maximal entanglement, i.e. $\mathcal C =1$, for $\theta=\pi$ and $p \approx 15.1 \, \text{MeV}$.
For a fixed momentum $p$, the outgoing electron has a single degree of freedom in $\theta$ such that $\bs{p}_4 = \bs{p}_4(\theta)$ (c.f. Sec~B of the Appendix).}
\label{plt1}
\end{figure}

\begin{figure}[h]
\centering
\includegraphics[width=\columnwidth]{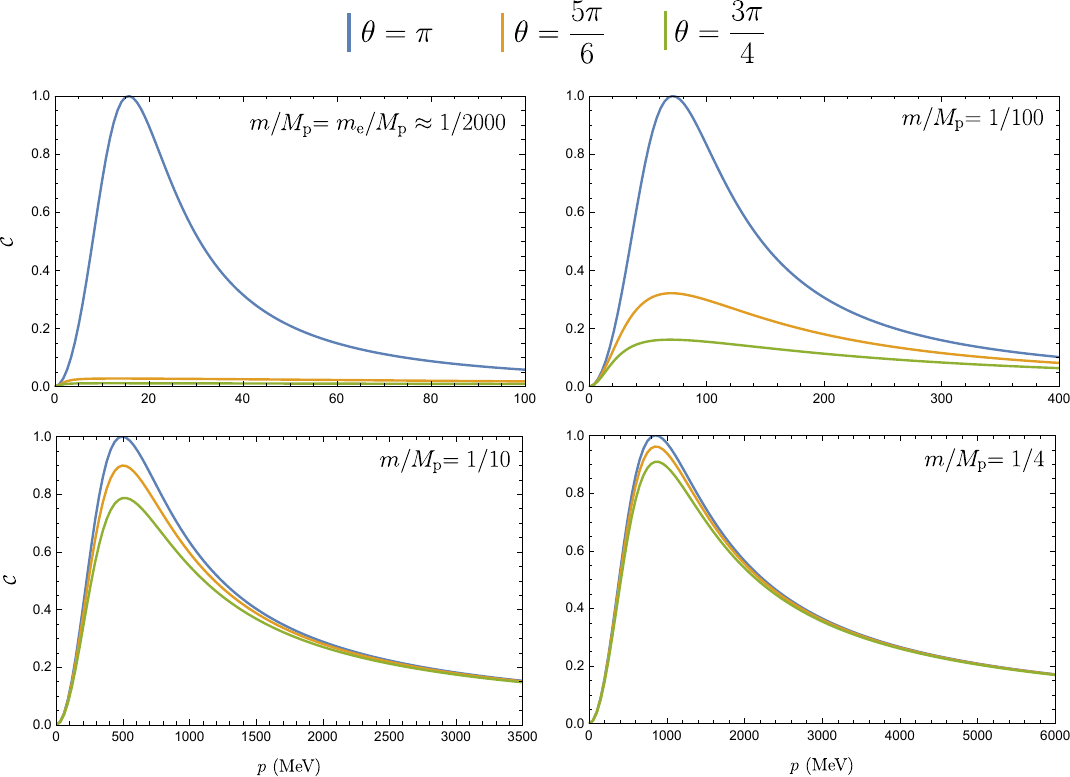}
\caption{Concurrence between the scattered particles qubit helicity degrees of freedom for the single scattering scenario. The target particle is fixed as proton with mass $M=M_\text{p}$, while the smaller particle is considered for four different possible masses. In each of the four cases, the concurrence is plotted for a set of values of the scattering angle $\theta$ to illustrate how the entanglement is more easily generated for higher values of the mass ratio $m/M_\text{p}$. This however comes at the expense of requiring higher momenta for the accelerated particle to achieve maximum entanglement.}
\label{plt4}
\end{figure}

Figure~\ref{plt1} illustrates the entanglement generation for different initial momentum $p$ and scattering angle $\theta$ for an initial state of the form $\ket{01}$, for the particular case of an electron scattering off a proton. One finds that, for most angles and momenta, the scattering does not generate any entanglement. 
However, it is interesting to note that for situations where the electron reflects off the proton, i.e. $\theta \approx \pi$, the particles become entangled, and can even reach maximal entanglement for $\theta = \pi $ and $p=\sqrt{\frac{M m}2} \left(1 - \sqrt{\frac{m/M}2} + \mathcal{O}(m/M) \right) \approx 15.1 \, \text{MeV}$, an energy which is not high enough to relevantly probe the proton's internal structure, so we ignore the contribution of the proton's form factor to the scattering amplitudes. The existence of an optimal momentum is actually expected on intuitive grounds: for very small or very large momenta, the projected particle barely interacts with the target particle, leading to small entanglement, so there must be a middle ground where the interaction is stronger, leading to the maximal possible generated entanglement. Nevertheless, the same process between particles with different mass ratios can accommodate a much larger $\theta$ variation. To see this, we can redo the concurrence plot of Figure~\ref{plt1} for different projectile particle masses, as is done in Figure~\ref{plt4}. It is evident that it becomes possible generate entanglement for larger angular apertures as the projectile particle's mass approaches that of the target particle's. This will come at the cost of requiring a larger momentum for the projectile particle.


\section{Consecutive Scatterings}

We are now interested in the situation where, after the first 2-particle scattering, one of the scattered particles interacts again with a third particle, without ever having its state collapsed. Assuming that enough time and space are given, the second scatter takes as initial states particles which are under the influence of the free Hamiltonian, thus one may consider again that the interaction is governed by the unitary operator of Eq.~(\ref{finalstate}). We can now apply this operator two consecutive times: one for the scattering of the first projectile particle with the target particle and another for the scattering of the second projectile particle with the recoiled target particle. Finally, since we wish to study the entanglement generated in the qubit degrees of freedom associated to the helicities, we will focus on the helicity component of the final state. The result is the following state (proof in Section A of the Appendix):
\eq{
\ket{\psi_h} = \ma{N}_3 \sum_{h_4,h_5,h_6} d_{h_4 h_5 h_6}(\bs{p}_4, \bs{p}_5, \bs{p}_6) \ket{h_4, h_5, h_6}
}
where $\bs{p}_4$, $\bs{p}_5$, $\bs{p}_6$ and $h_4$, $h_5$, $h_6$ are the momenta and helicities of the target, first and second projectile particles, respectively, and
\ea{
& d_{h_4 h_5 h_6}(\bs{p}_4, \bs{p}_5, \bs{p}_6) = \nonumber \\
& \hspace{4mm} \sum_{r_1}  \ma{M}(\bs{p}_4, h_4; \bs{p}_6, h_6 | \bs{p}_5 - \bs{p}_1 - \bs{p}_2, r_1; \bs{p}_3, h_3) \, \nonumber \\
& \hspace{9mm} \times \ma{M}(\bs{p}_5 - \bs{p}_1 - \bs{p}_2, r_1; \bs{p}_5, h_5 | \bs{p}_1, h_1; \bs{p}_2, h_2)
}
are the components of the state, with
\ea{
& \ma{N}_3 =  - \frac{(2\pi)^5}{2E_{\bs{p}_5 - \bs{p}_1 - \bs{p}_2}}  \de\left(E_{\bs{p}_5 - \bs{p}_1 - \bs{p}_2} + E_{\bs{p}_5} - E_{\bs{p}_1} - E_{\bs{p}_2}\right) \nonumber \\
& \hspace{10mm} \times \de\left(E_{\bs{p}_5 - \bs{p}_1 - \bs{p}_2}+E_{\bs{p}_3}-E_{\bs{p}_4}-E_{\bs{p}_6}\right) \nonumber \\
& \hspace{10mm} \times \de^{(3)}\left(\bs{p}_1+\bs{p}_2+\bs{p}_3-\bs{p}_4-\bs{p}_5-\bs{p}_6\right)
}
being formally a constant which enforces conservation of momentum and energy over the set of scatterings that occur by means of Dirac deltas. To study the entanglement of the system one first needs to construct the density matrix
\eq{
\rho_{t,p,p} = \frac{\ket{\psi_h}\bra{\psi_h}}{\textrm{Tr}[\ket{\psi_h}\bra{\psi_h}]}\,,
}
whose normalization eliminates the cumbersome $\ma{N}_3$ factor. The amount of bipartite entanglement generated between the two projectile particles is obtained by using the reduced density matrix derived by tracing out the target particle qubit $t$, resulting in the density matrix (proof in Section A of the Appendix)
\ea{
\rho_{p,p} & = \textrm{Tr}_{t}[\rho_{t,p,p}] \nonumber \\
& = \sum_{h_5,h_6,r_5,r_6} \rho_{h_5 h_6 r_5 r_6} \ket{h_5, h_6}  \bra{r_5, r_6}
}
where we have the reduced density matrix components
\eq{
\rho_{h_5 h_6 r_5 r_6} = \frac{\sum_{\si} d_{\si h_5 h_6}(\bs{p}_4, \bs{p}_5, \bs{p}_6)d^{*}_{\si r_5 r_6}(\bs{p}_4, \bs{p}_5, \bs{p}_6)}{\sum_{h_4,h_5,h_6} |d_{h_4 h_5 h_6}(\bs{p}_4, \bs{p}_5, \bs{p}_6)|^2}\,. \label{rhoee}
}
To calculate the entanglement between the two projectiles' helicities we find the concurrence $\mathcal{C} \, [p,p]$ by using Eq.~(\ref{eqconcurrence}) for the reduced density matrix in Eq.~(\ref{rhoee}).

\begin{figure}[h]{}
\centering
\includegraphics[width=\columnwidth]{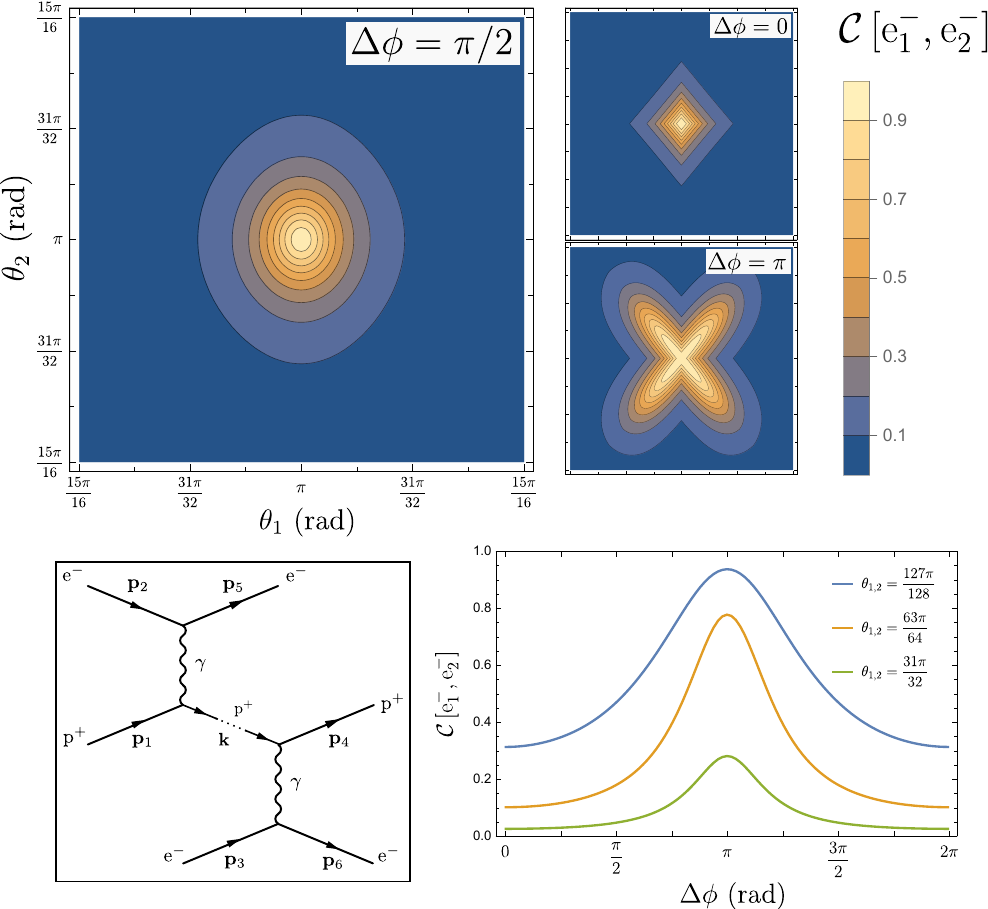}
\caption{Bottom left: Feynman diagrams for two consecutive scatterings between two electrons and a proton. The dashed line represents a scattered proton external line for both diagrams. Top half and bottom right: Concurrence generated from scattering two electrons with momenta $p\approx 47.16 \, \text{MeV}$ off a proton at rest in the laboratory frame. 
Choosing for the initial state a polarized state where both electrons have a helicity orthogonal to the proton's starting helicity maximises the entnanglement generated. The relative positions in the cone are also relevant.
For a fixed momentum $p$, the outgoing electrons have a total of three degrees of freedom $\theta_1$, $\theta_2$ and $\Delta \phi$, such that $\bs{p}_5=\bs{p}_5(\theta_1)$ and $\bs{p}_6=\bs{p}_6(\theta_2,\Delta \phi)$ (c.f. Sec.~B of the Appendix). Entanglement is maximised for $\theta_{1,2}=\pi$, $\Delta \phi=\pi$, and $p\approx 47.16 \, \text{MeV}$, but fails to reach maximal value between the two electrons.}
\label{plt2b}
\end{figure}

We now wish to apply the precedent reasoning to the scenario where two projectiles originating from the same source, i.e. both described by the four momentum $p_1 = (\sqrt{m^2+p^2}, 0,0,p)$, scatter at different instants off a target which started at rest, as illustrated in the diagram in Figure~\ref{plt2b}.
It can be shown that only three degrees of freedom are associated to the two outgoing projectiles (c.f. Sect.~B of the Appendix): the polar angle $\theta_1$ of the first scattered projectile, the polar angle $\theta_2$ of the second scattered projectile, and the azimuth angle difference between the two, denoted by $\Delta \phi$. As reported in the previous section, entanglement is maximised for $\theta$ close to $\pi$, so it is expected that the two projectiles are more entangled close to this value as well. Figure~\ref{plt2b} evidently displays this behavior, for a fixed momentum $p\approx 47.16\,\text{MeV}$ which maximises the concurrence for the particular case of electrons being fired at a proton.
It is interesting to note how the degree of freedom $\Delta \phi$ affects the concurrence in Figure~\ref{plt2b}.
If we consider a fixed value for both $\theta_1$ and $\theta_2$, we can interpret the two electrons as being reflected along the surface of a cone centered around the $z$-axis, such that $\Delta \phi$ indicates how distant the two electrons lie on the cone.
It is clear that, the more distant apart the electrons are in the surface of the cone, the more entanglement is generated, reaching a maximum when they are on opposite sides of the cone with $\Delta \phi= \pi$.

The electron momentum maximising the entanglement in this scenario is larger than the one required in the single scattering scenario of Figure~\ref{plt1}. This is due to the fact that the proton from the first scattering acquires a momentum of roughly $2p$ after the initial scatter, where $\theta_1$ is close to $\pi$. Since the scattered proton is moving away from the second electron, the relative momentum between the two is thus smaller than in the first scattering so the electrons momentum needs to be larger to compensate for this difference, in order to achieve the optimal relative momentum derived in Sec.~III.

\begin{figure}[h!]
\centering
\includegraphics[width=\columnwidth]{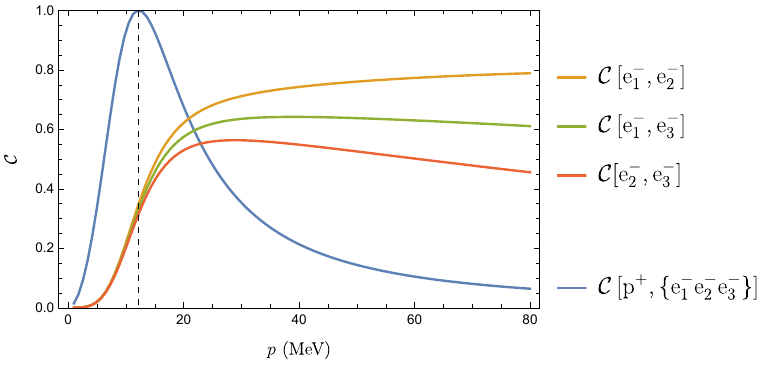}
\caption{Concurrence between substystems resulting from three reflected collisions. The proton can be maximally entangled with the subsystem of three electrons, which describes a qudit of $2^3$ levels, at a relatively low energy of $p\approx 12.06\, \text{MeV}$.}
\label{plt3i}
\end{figure}

To have an intuition of how the entanglement will be distributed with more scatterings, one can calculate the case of three scatterings. The latter calculations are straightforwardly generalizable from the derivations in Secs.~A and B of the Appendix. Although there isn't a single measure to quantify the entanglement present in the resulting 4-quibt state, one can calculate the entanglement between the relevant bipartitions, namely between all pairs of electrons and between the proton and all other electrons. This is represented in Figure~\ref{plt3i}.
It becomes evident that, as more collisions occur, the amount of entanglement generated between the latest pairs of electrons tends to decrease. This is expected since the proton keeps moving further away with more momentum as more collisions take place and so each consecutive electron carries less relative momentum with respect to the proton. In addition to this, there is an additional physical restriction at play, namely the monogamy of entanglement \cite{PhysRevA.61.052306}. The latter result essencially states that, in order for two qubits A and B to be maximally entangled, they must not be entangled with a third subsystem C. Since all incolved particles are entangled in this case, no pair of particles can ever be maximally entangled. In fact, only the bipartition proton/electrons can actually reach maximal entanglement.


\section{Conclusions}

In this work we studied how the consecutive scatterings of particles (originating from the same source) with a central target particle can be used to generate a multipartite entangled state between all particles. In particular, we showed that the amount of bipartite entanglement entanglement between pairs of particles, as quantified by the concurrence, is a balance between two factors. On one hand, the entanglement increases the closer the projectile particle is to total reflection as well as how close its mass is to the target particle's. On the other hand, the entanglement decreases with the number of collisions while at the same time making it harder for consecutive projectile particles to hit the target particle. For the specific case of electrons scattering off a proton, it is clear that any appreciable entanglement is concentrated on a small solid angle centered around the total reflected direction, with larger angular apertures for larger mass ratios. It was also shown that there is always a momentum for which the generated entanglement achieves a maximum value, which can be associated to an optimal point of interaction lying between the extremes of small and large momenta, where the projectile and target particles barely interact.

As a method to generate multipartite quantum states of massive particles, the consecutive scatterings of particles from a heavier target particle is conceptually much simpler than most currently used methods. Nevertheless, there are two clear bottlenecks to be addressed for experimental implementations. Firstly, for small mass ratios (the easiest to implement) one would need to focus on totally reflected particles, whose direction collides with the newly projected particles. Secondly, the heavier target particle would need to be as static as possible for the projectiled particles to hit it, and the method used to keep it still could influence the projected particles in non-trivial ways. While the second diffculty would evidently be harder to address, a solution to the first one might be to use magnetic fields to alter the trajectory of the projectile particles, which would not change their energies. One would need, however, to take into account the innevitable effects of spin precession in the helicity qubit degrees of freedom.


\section{Acknowledgements}

The authors thank the support from Funda\c{c}\~{a}o para a Ci\^{e}ncia e a Tecnologia (Portugal), namely through project CEECIND/02474/2018 and project EXPL/FIS-PAR/1604/2021 QEntHEP - Quantum Entanglement in High Energy Physics.


\clearpage
\pagebreak
\newpage

\widetext
\begin{center}
\textbf{\large Supplemental Material for ``Multipartite Entanglement from Consecutive Scatterings''}
\end{center}
\setcounter{equation}{0}
\setcounter{figure}{0}
\setcounter{table}{0}
\setcounter{page}{1}
\makeatletter
\renewcommand{\theequation}{A\arabic{equation}}
\renewcommand{\thefigure}{A\arabic{figure}}
\renewcommand{\bibnumfmt}[1]{[A#1]}
\renewcommand{\citenumfont}[1]{A#1}

\section*{Section A: Explicit State Construction for two projectile-target Scatterings}

Consider an initial state composed by a target with 4-momentum $\bs{p}_1$ and helicity $h_1$ and two projectiles with momenta $\bs{p}_2$, $\bs{p}_3$ and helicities $h_2$ and $h_3$. Consider as well the unitary matrix defined by
\eq{
\hat{U}_{i,j} \ket{\bs{p}_1, r_1; \ldots; \bs{p}_n, r_n} \equiv \left(\hat{U} \ket{\bs{p}_i, r_i; \bs{p}_j, r_j}\right) \otimes \ket{\bs{p}_1, r_1; \ldots; \bs{p}_{i-1}, r_{i-1}; \bs{p}_{i+1}, r_{i+1}; \ldots; \bs{p}_{j-1}, r_{j-1}; \bs{p}_{j+1}, r_{j+1}; \ldots; \bs{p}_n, r_n}
}
i.e. it applies the unitary evolution operator $\hat{U}$, defined by Eq.~(\ref{evolutionU}), on the two particles indexed by $i$ and $j$ (the positions relative to the initial ket). Using this notation, the scattering of a target particle with one projectile particle, followed by another scattering of the same target with another projectile, will result in a final state of the form
\eq{
\ket{\psi(t_f)} = \hat{U}_{1,3} \hat{U}_{1,2} \ket{\bs{p}_1, r_1; \bs{p}_2, r_2; \bs{p}_3, r_3}\,.
}
To find the explicit form of this state, one may note the specific form of Eq.~(\ref{genFinalState}) for 2-particle scatterings, namely
\ea{\label{2scatt}
\hat{U}_{1,2} \ket{\bs{p}_1, h_1; \bs{p}_2, h_2} & = \ket{\bs{p}_1, h_1; \bs{p}_2, h_2} + i \int \frac{d^3 \bs{q}_1}{(2\pi)^3 2E_{\bs{q}_1}} \frac{d^3 \bs{q}_2}{(2\pi)^3 2E_{\bs{q}_2}} (2\pi)^4 \de^{(4)}\left(q_1+q_2-p_1-p_2\right) \nonumber \\
& \hspace{10mm} \times \sum_{r_1, r_3}  \ma{M}(\bs{q}_1, r_1; \bs{q}_2, r_2 | \bs{p}_1, h_1; \bs{p}_2, h_2) \ket{\bs{q}_1, r_1; \bs{q}_2, r_2}\,.
}
We can now start by considering the first scattering only, whose resulting state is
\ea{
\hat{U}_{1,2} \ket{\bs{p}_1, h_1; \bs{p}_2, h_2; \bs{p}_3, h_3} & = \ket{\bs{p}_1, h_1; \bs{p}_2, h_2; \bs{p}_3, h_3} +  i \int \frac{d^3 \bs{q}_1}{(2\pi)^3 2E_{\bs{q}_1}} \frac{d^3 \bs{q}_2}{(2\pi)^3 2E_{\bs{q}_2}} (2\pi)^4 \de^{(4)}\left(q_1 + q_2 - p_1 - p_2\right) \nonumber \\
& \hspace{10mm} \times \sum_{r_1, r_2}  \ma{M}(\bs{q}_1, r_1; \bs{q}_2, r_2 | \bs{p}_1, h_1; \bs{p}_2, h_2) \ket{\bs{q}_1, r_1; \bs{q}_2, r_2; \bs{p}_3, h_3}\,.
}
The second target-projectile scattering, encoded by $\hat{U}_{1,3}$, results in the state
\ea{
\hat{U}_{1,3} \hat{U}_{1,2} \ket{\bs{p}_1, h_1; \bs{p}_2, h_2; \bs{p}_3, h_3} & = \nonumber \\
& \hspace{-40mm} =  \hat{U}_{1,3}\ket{\bs{p}_1, h_1; \bs{p}_3, h_3}\otimes \ket{\bs{p}_2, h_2} +  i \int \frac{d^3 \bs{q}_1}{(2\pi)^3 2E_{\bs{q}_1}} \frac{d^3 \bs{q}_2}{(2\pi)^3 2E_{\bs{q}_2}} (2\pi)^4 \de^{(4)}\left(q_1 + q_2 - p_1 - p_2\right) \nonumber \\
& \hspace{10mm} \times \sum_{r_1, r_2}  \ma{M}(\bs{q}_1, r_1; \bs{q}_2, r_2 | \bs{p}_1, h_1; \bs{p}_2, h_2) \hat{U}_{1,3} \ket{\bs{q}_1, r_1; \bs{p}_3, h_3} \ket{\bs{q}_2, r_2} \nonumber \\
& \hspace{-40mm} = \ket{\bs{p}_1, h_1; \bs{p}_2, h_2; \bs{p}_3, h_3} +  i \int \frac{d^3 \bs{q}_1}{(2\pi)^3 2E_{\bs{q}_1}} \frac{d^3 \bs{q}_3}{(2\pi)^3 2E_{\bs{q}_3}} (2\pi)^4 \de^{(4)}\left(q_1 + q_3 - p_1 - p_3\right) \nonumber \\
& \hspace{10mm} \times \sum_{r_1, r_2}  \ma{M}(\bs{q}_1, r_1; \bs{q}_3, r_3 | \bs{p}_1, h_1; \bs{p}_3, h_3) \ket{\bs{q}_1, r_1; \bs{p}_2, h_2; \bs{q}_3, r_3} \nonumber \\
& \hspace{-35mm} +  i \int \frac{d^3 \bs{q}_1}{(2\pi)^3 2E_{\bs{q}_1}} \frac{d^3 \bs{q}_2}{(2\pi)^3 2E_{\bs{q}_2}} (2\pi)^4 \de^{(4)}\left(q_1 + q_2 - p_1 - p_2\right)  \sum_{r_1, r_2}  \ma{M}(\bs{q}_1, r_1; \bs{q}_2, r_2 | \bs{p}_1, h_1; \bs{p}_2, h_2) \times \nonumber \\
& \hspace{-40mm} \times \biggl(\ket{\bs{q}_1, r_1; \bs{p}_3, h_3} + i \int \frac{d^3 \bs{k}_1}{(2\pi)^3 2E_{\bs{k}_1}}\frac{d^3 \bs{k}_3}{(2\pi)^3 2E_{\bs{k}_3}} (2\pi)^4 \de^{(4)}\left(q_1+p_3-k_1-k_3\right) \nonumber \\
& \hspace{20mm} \times \sum_{l_1, l_3}  \ma{M}(\bs{k}_1, l_1; \bs{k}_3, l_3 | \bs{q}_1, r_1; \bs{p}_3, h_3) \ket{\bs{k}_1, l_1; \bs{k}_3, l_3}\biggr) \ket{\bs{q}_2, r_2} \nonumber \\
}
\ea{
& = \ket{\bs{p}_1, h_1; \bs{p}_2, h_2; \bs{p}_3, h_3} + \nonumber \\
& \hspace{-5mm} + i \int \frac{d^3 \bs{q}_1}{(2\pi)^3 2E_{\bs{q}_1}} \frac{d^3 \bs{q}_3}{(2\pi)^3 2E_{\bs{q}_3}} (2\pi)^4 \de^{(4)}\left(q_1 + q_3 - p_1 - p_3\right) \sum_{r_1, r_2}  \ma{M}(\bs{q}_1, r_1; \bs{q}_3, r_3 | \bs{p}_1, h_1; \bs{p}_3, h_3) \ket{\bs{q}_1, r_1; \bs{p}_2, h_2; \bs{q}_3, r_3} \nonumber \\
& \hspace{-5mm} +  i \int \frac{d^3 \bs{q}_1}{(2\pi)^3 2E_{\bs{q}_1}} \frac{d^3 \bs{q}_2}{(2\pi)^3 2E_{\bs{q}_2}} (2\pi)^4 \de^{(4)}\left(q_1 + q_2 - p_1 - p_2\right)  \sum_{r_1, r_2}  \ma{M}(\bs{q}_1, r_1; \bs{q}_2, r_2 | \bs{p}_1, h_1; \bs{p}_2, h_2) \ket{\bs{q}_1, r_1; \bs{q}_2, r_2; \bs{p}_3, h_3} \nonumber \\
& \hspace{-5mm} + i^2 \int \frac{d^3 \bs{q}_1}{(2\pi)^3 2E_{\bs{q}_1}} \frac{d^3 \bs{q}_2}{(2\pi)^3 2E_{\bs{q}_2}} \frac{d^3 \bs{k}_1}{(2\pi)^3 2E_{\bs{k}_1}}\frac{d^3 \bs{k}_3}{(2\pi)^3 2E_{\bs{k}_3}} (2\pi)^8 \de^{(4)}\left(q_1 + q_2 - p_1 - p_2\right) \de^{(4)}\left(q_1+p_3-k_1-k_3\right) \nonumber \\
& \hspace{-5mm} \times \sum_{r_1, r_2, l_1, l_3}  \ma{M}(\bs{k}_1, l_1; \bs{k}_3, l_3 | \bs{q}_1, r_1; \bs{p}_3, h_3) \, \ma{M}(\bs{q}_1, r_1; \bs{q}_2, r_2 | \bs{p}_1, h_1; \bs{p}_2, h_2)  \ket{\bs{k}_1, l_1; \bs{q}_2, r_2; \bs{k}_3, l_3} \,.}
The final state is thus a superposition of 4 states: one where no scattering occured, two others where one of the scatterings didn't occur and a final one where all scatterings happened. Using now the normalization relation
\eq{
\braket{\bs{k},r | \bs{p}, s} = (2\pi)^3 \de_{rs} \de^{(3)}(\bs{k}-\bs{p}) 2 E_{\bs{p}}
}
we can find the probability amplitude of obtaing as final state a target with momentum $\bs{p}_4$ and helicity $h_4$ as well as two projectiles with momenta $\bs{p}_5$ and $\bs{p}_6$ and helicities $h_5$ and $h_6$. This comes out as
\ea{
\bra{\bs{p}_4, h_4; \bs{p}_5, h_5; \bs{p}_6, h_6}\hat{U}_{1,3} \hat{U}_{1,2} \ket{\bs{p}_1, h_1; \bs{p}_2, h_2; \bs{p}_3, h_3} & = \nonumber \\
& \hspace{-55mm} = i^2 \int \frac{d^3 \bs{q}_1}{(2\pi)^3 2E_{\bs{q}_1}} \frac{d^3 \bs{q}_2}{(2\pi)^3 2E_{\bs{q}_2}} \frac{d^3 \bs{k}_1}{(2\pi)^3 2E_{\bs{k}_1}}\frac{d^3 \bs{k}_3}{(2\pi)^3 2E_{\bs{k}_3}} (2\pi)^8 \de^{(4)}\left(q_1 + q_2 - p_1 - p_2\right) \de^{(4)}\left(q_1+p_3-k_1-k_3\right) \nonumber \\
& \hspace{-45mm} \times (2\pi)^3 \de_{h_4 l_1} \de^{(3)}(\bs{k}_1-\bs{p}_4) 2 E_{\bs{k}_1} (2\pi)^3 \de_{h_5 r_2} \de^{(3)}(\bs{q}_2-\bs{p}_5) 2 E_{\bs{q}_2} (2\pi)^3 \de_{h_6 l_3} \de^{(3)}(\bs{k}_3-\bs{p}_6) 2 E_{\bs{k}_3} \nonumber \\
& \hspace{-45mm} \times \sum_{r_1, r_2, l_1, l_3}  \ma{M}(\bs{k}_1, l_1; \bs{k}_3, l_3 | \bs{q}_1, r_1; \bs{p}_3, h_3) \, \ma{M}(\bs{q}_1, r_1; \bs{q}_2, r_2 | \bs{p}_1, h_1; \bs{p}_2, h_2) \nonumber \\
& \hspace{-55mm} = - \int \frac{d^3 \bs{q}_1}{2E_{\bs{q}_1}} (2\pi)^5 \de^{(4)}\left(q_1 + p_5 - p_1 - p_2\right) \de^{(4)}\left(q_1+p_3-p_4-p_6\right) \nonumber \\
& \hspace{-45mm} \times \sum_{r_1}  \ma{M}(\bs{p}_4, h_4; \bs{p}_6, h_6 | \bs{q}_1, r_1; \bs{p}_3, h_3) \, \ma{M}(\bs{q}_1, r_1; \bs{p}_5, h_5 | \bs{p}_1, h_1; \bs{p}_2, h_2) \nonumber \\
& \hspace{-55mm} = - \frac{(2\pi)^5}{2E_{\bs{p}_5 - \bs{p}_1 - \bs{p}_2}}  \de\left(E_{\bs{p}_5 - \bs{p}_1 - \bs{p}_2} + E_{\bs{p}_5} - E_{\bs{p}_1} - E_{\bs{p}_2}\right) \nonumber \\
& \hspace{-45mm} \de\left(E_{\bs{p}_5 - \bs{p}_1 - \bs{p}_2}+E_{\bs{p}_3}-E_{\bs{p}_4}-E_{\bs{p}_6}\right) \de^{(3)}\left(\bs{p}_1+\bs{p}_2+\bs{p}_3-\bs{p}_5-\bs{p}_4-\bs{p}_6\right) \nonumber \\
& \hspace{-45mm} \times \sum_{r_1}  \ma{M}(\bs{p}_4, h_4; \bs{p}_6, h_6 | \bs{p}_5 - \bs{p}_1 - \bs{p}_2, r_1; \bs{p}_3, h_3) \, \ma{M}(\bs{p}_5 - \bs{p}_1 - \bs{p}_2, r_1; \bs{p}_5, h_5 | \bs{p}_1, h_1; \bs{p}_2, h_2) \nonumber \\
& \hspace{-55mm} \equiv \ma{N}_3 d_{h_4 h_5 h_6}(\bs{p}_4, \bs{p}_5, \bs{p}_6)
}
where
\eq{
d_{h_4 h_5 h_6}(\bs{p}_4, \bs{p}_5, \bs{p}_6) = \sum_{r_1}  \ma{M}(\bs{p}_4, h_4; \bs{p}_6, h_6 | \bs{p}_5 - \bs{p}_1 - \bs{p}_2, r_1; \bs{p}_3, h_3) \, \ma{M}(\bs{p}_5 - \bs{p}_1 - \bs{p}_2, r_1; \bs{p}_5, h_5 | \bs{p}_1, h_1; \bs{p}_2, h_2)
}
and we formally take
\eq{
\ma{N}_3 =  - \frac{(2\pi)^5}{2E_{\bs{p}_5 - \bs{p}_1 - \bs{p}_2}}  \de\left(E_{\bs{p}_5 - \bs{p}_1 - \bs{p}_2} + E_{\bs{p}_5} - E_{\bs{p}_1} - E_{\bs{p}_2}\right) \de\left(E_{\bs{p}_5 - \bs{p}_1 - \bs{p}_2}+E_{\bs{p}_3}-E_{\bs{p}_4}-E_{\bs{p}_6}\right)
}
to be a constant. One sees that $\ma{N}_3$ naturally contains conservation of energy and momentum over the set of scatterings that occur.

We are now interested in analyzing the entanglement created in the helicity degrees of freedom of the final outgoing state in the case where the two scatterings occur. This amounts to looking at the helicity component of the final state, which will have the form
\eq{
\ket{\psi_h} = \ma{N}_3 \sum_{h_4,h_5,h_6} d_{h_4 h_5 h_6}(\bs{p}_4, \bs{p}_5, \bs{p}_6) \ket{h_4, h_5, h_6}\,.
}
One may now trace away the target subsystem and check if the remaing parts, composed by the two projectiles, are entangled. In order to do that, one must first build the density matrix. Noting that $\ma{N}_3$ only involves real quantities, we may consider that $\textrm{Tr}[\ket{\psi_h}\bra{\psi_h}]=\ma{N}^2 \sum_{h_4,h_5,h_6} |d_{h_4 h_5 h_6}(\bs{p}_4, \bs{p}_5, \bs{p}_6)|^2$, so that a normalized density matrix for the final state is
\ea{
\rho_{t,p,p} & = \frac{\ket{\psi_h}\bra{\psi_h}}{\textrm{Tr}[\ket{\psi_h}\bra{\psi_h}]} \nonumber \\
& = \frac{\sum_{h_4,h_5,h_6,r_4,r_5,r_6} d_{h_4 h_5 h_6}(\bs{p}_4, \bs{p}_5, \bs{p}_6)d^{*}_{r_4 r_5 r_6}(\bs{p}_4, \bs{p}_5, \bs{p}_6) \ket{h_4, h_5, h_6}  \bra{r_4,r_5, r_6}}{\sum_{h_4,h_5,h_6} |d_{h_4 h_5 h_6}(\bs{p}_4, \bs{p}_5, \bs{p}_6)|^2}
} 
where we used the indexes $t$ and $p$ to refer to target and projectile subsystems, respectively. Note that the factor of $\ma{N}_3$ has formally been cancelled out. Tracing out the target subsystem results in the system composed by the two projectiles, whose entanglement properties we can then investigate.

We must now perform the partial trace of the helicity qubits related to the proton, which amounts to applying the formula
\eq{
\textrm{Tr}_{t}[\rho] =  \sum_{\si} (I_r \otimes \bra{\si}_{t}) \rho (I_r \otimes \ket{\si}_{t})
}
where $I_r$ represents the identity acting on the remaining subsystems. Defining for simplicity
\eq{
\ma{N}_{\rho} = \sum_{h_4,h_5,h_6} |d_{h_4 h_5 h_6}(\bs{p}_4, \bs{p}_5, \bs{p}_6)|^2
}
we obtain the partially traced density matrix
\ea{
\rho_{p,p} & = \textrm{Tr}_{t}[\rho_{t,p,p}] \nonumber \\
& = \ma{N}^{-1}_{\rho} \sum_{\si,h_4,h_5,h_6,r_4,r_5,r_6} d_{h_4 h_5 h_6}(\bs{p}_4, \bs{p}_5, \bs{p}_6)d^{*}_{r_4 r_5 r_6}(\bs{p}_4, \bs{p}_5, \bs{p}_6) \ket{h_5, h_6}  \bra{r_5, r_6} \nonumber \\
& = \sum_{h_5,h_6,r_5,r_6} \rho_{h_5 h_6 r_5 r_6} \ket{h_5, h_6}  \bra{r_5, r_6}\,,
}
where we have the reduced density matrix components
\eq{
\rho_{h_5 h_6 r_5 r_6} = \ma{N}^{-1}_{\rho} \left(\sum_{\si} d_{\si h_5 h_6}(\bs{p}_4, \bs{p}_5, \bs{p}_6)d^{*}_{\si r_5 r_6}(\bs{p}_4, \bs{p}_5, \bs{p}_6) \right)\,.
}
The matrix $\rho_{p,p}$ has trace equal to 1, as can be straightforwardly checked and will in general be mixed, i.e. $\textrm{Tr}[\rho^2_{e,e}]<1$.


\section*{Section B: Kinematics for consecutive collisions}

\begin{figure}[h]
\centering
\includegraphics[width=.5\columnwidth]{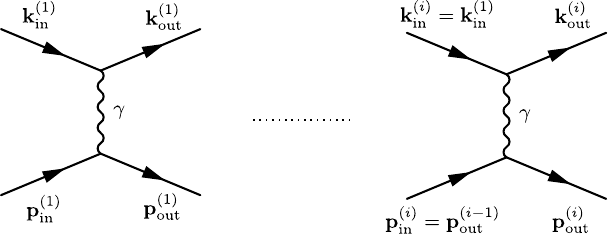}
\caption{Diagram depicting the consecutive scatterings. The diagrams represent the tree-level QED interaction between two distinguishable types of particles denoted by the 3-momenta $\bs{k}$ and \bs{p}, with rest-masses $m$ and $M$ respectively. The superscripts in the 3-momenta indicate how many interactions occurred up to that point, while the subscripts indicate whether the particle is incoming or outgoing. The 3-momenta $\bs{p}$ indicate the target particle which started off at rest, while $\bs{k}$ is the lighter particle which is accelerated into the target. For each interaction, a different particle $\bs{k}$, with the same energy and direction as the first one, is shot at the same target particle $\bs{p}$, such that the in the $i$-th interaction, $\bs{k}^{(i)}_\text{in} = \bs{k}^{(1)}_\text{in}$, while $\bs{p}^{(i)}_\text{in} = \bs{p}^{(i-1)}_\text{out}$.}
\label{pltappendixb}
\end{figure}

Here we demonstrate the steps required to reproduce in all generality the kinematics for the consecutive collisions that take place in the main body of the paper. For the sake of generalisation, we will employ a different notation for the ingoing and outgoing momenta, as depicted in Fig.~\ref{pltappendixb}.

Let $\bs{p}^{(1)}_\text{in}$ be the 3-momentum of the target particle which starts of at rest and has mass $M$. The superscript $(1)$ indicates that we are referring to the first scattering that particle $\bs{p}$ is experiencing, while the subscript 'in' indicates that it is an incoming particle.
The other incoming particle in the first scattering will be denoted by $\bs{k}^{(1)}_\text{in}$, which is the particle that will be accelerated towards the target particle with linear momentum $p$ and mass $m$.
After a single scattering, the outgoing 3-momenta are $\bs{p}^{(1)}_\text{out}$ and $\bs{k}^{(1)}_\text{out}$.
From here, we want to consider additional copies of the particle $\bs{k}^{(1)}_\text{in}$ to consecutively scatter against the same target particle, so for the second collision we will have $\bs{k}^{(2)}_\text{in} = \bs{k}^{(1)}_\text{in}$ scattering off of $\bs{p}^{(2)}_\text{in}=\bs{p}^{(1)}_\text{out}$, and so on for any number of intended collisions.
With this we set up the first conditions for our system of consecutive collisions, which will allows to recursively compute the kinematics in the $i$-th interaction. In particular, we have:
\begin{align}
\bs{p}^{(1)}_\text{in} &= (0,0,0),\label{consecutivestart} \\
\bs{k}^{(1)}_\text{in} &= (0,0,p), \\
\bs{p}^{(i)}_\text{in} &= \bs{p}^{(i-1)}_\text{out}, \\
\bs{k}^{(i)}_\text{in} &= \bs{k}^{(1)}_\text{in}.\label{consecutivefinish}
\end{align}

As we will see, each interaction will result in an additional two degrees of freedom, which we will take as the angular coordinates for the outgoing particle $\bs{k}^{(i)}_\text{out}$, namely $\theta^{(i)}_\bs{k}$ and $\phi^{(i)}_\bs{k}$. Due to spherical symmetry, one can eliminate a single degree of freedom $\phi^{(i)}_\bs{k}$, which we choose to be $\phi^{(1)}_\bs{k}$ in this work, although any other $\phi^{(i)}_\bs{k}$ is possible. In order to find the values of $\bs{p}^{(i)}_\text{out}$ and $\bs{k}^{(i)}_\text{out}$ as functions of $\theta^{(i)}_\bs{k}$ and $\phi^{(i)}_\bs{k}$, we only need to guarantee that these can be recursively calculated from $\bs{p}^{(i-1)}_\text{out} = \bs{p}^{(i)}_\text{in}$.

Let $E_{\bs{k}}$ and $E_{\bs{p}}$ denote the energies the particles with momenta $\bs{k}$ and $\bs{p}$ respectively.
From the conservation of 4-momentum leading to the equality $\left(k^{(i)}_\text{in} - k^{(i)}_\text{out}\right)^2 = \left(p^{(i)}_\text{out} - p^{(i)}_\text{in}\right)^2$ and from the conservation of energy $E_{\bs{p}^{(i)}_\text{out}} = E_{\bs{k}^{(i)}_\text{in}} + E_{\bs{p}^{(i)}_\text{in}} - E_{\bs{k}^{(i)}_\text{out}}$, one can write
\begin{equation}\label{modkequation}
\sqrt{\left|\bs{k}^{(i)}_\text{out}\right|^2 + m^2} = a + b \left|\bs{k}^{(i)}_\text{out}\right|,
\end{equation}
with
\begin{align}
a &\equiv E_{\bs{p}^{(i)}_\text{in}} - \frac{M^2-m^2 + \bs{p}^{(i)}_\text{in} \cdot \left( \bs{p}^{(i)}_\text{in}+\bs{k}^{(i)}_\text{in} \right) }{E_{\bs{k}^{(i)}_\text{in}} + E_{\bs{p}^{(i)}_\text{in}}},\\
b &\equiv \frac{\vec{e}(\theta^{(i)}_\bs{k},\phi^{(i)}_\bs{k}) \cdot \left(\bs{p}^{(i)}_\text{in}+\bs{k}^{(i)}_\text{in}\right) }{E_{\bs{k}^{(i)}_\text{in}} + E_{\bs{p}^{(i)}_\text{in}}},
\end{align}
where $\vec{e}(\theta^{(i)}_\bs{k},\phi^{(i)}_\bs{k})$ is the unit vector for $\bs{k}^{(i)}_\text{out} = \left|\bs{k}^{(i)}_\text{out}\right| \vec{e}(\theta^{(i)}_\bs{k},\phi^{(i)}_\bs{k})$. Equation~(\ref{modkequation}) is solvable for $\left|\bs{k}^{(i)}_\text{out}\right|$ with a single positive solution
\begin{equation}\label{solvedk}
\left|\bs{k}^{(i)}_\text{out}\right| = \frac{a\,b + \sqrt{a^2 + m^2 (b^2-1)}}{b^2-1},
\end{equation}
and, so long as we know $\bs{p}^{(i)}_\text{in} = \bs{p}^{(i-1)}_\text{out}$, we can explicitly write Eq.~(\ref{solvedk}), which itself is enough to fully describe $\bs{k}^{(i)}_\text{out}$ as well as $\bs{p}^{(i)}_\text{out}$ (through 3-momentum conservation).

Finally, when calculating the scattering amplitudes it can be useful to define the angular coordinates $(\theta^{(i)}_\bs{p},\phi^{(i)}_\bs{p})$ associated to the direction of $\bs{p}^{(i)}_\text{out}$, to be used for the spinors in the helicity basis. These are straightforward to derive from $\bs{p}^{(i)}_\text{out}$. Let us consider the first interaction as an example. This case is simpler since $\bs{p}^{(1)}_\text{in} = (0,0,0)$, as well as by the fact that we can fix $\phi^{(1)}_\bs{k} = 0$ from spherical symmetry. As such, Eq.~(\ref{solvedk}) reads
\begin{equation}
\left|\bs{k}^{(1)}_\text{out}\right| = p \, \frac{ \left(m^2+M\sqrt{p^2+m^2} \right)\cos \left(\theta^{(1)}_\bs{k}\right) + \left( M + \sqrt{p^2+m^2} \right) \sqrt{M^2 - m^2 \sin^2 \left(\theta^{(1)}_\bs{k}\right) } }{\left( M + \sqrt{p^2+m^2} \right)^2 - p^2\cos^2 \left(\theta^{(1)}_\bs{k}\right)}.
\end{equation}
The form of $\bs{k}^{(1)}_\text{out}$ leads to a solution for $\bs{p}^{(1)}_\text{out}$, which in turn leads to a recursive solution for any $i$-th interaction as given by Eqs.~(\ref{consecutivestart})-({\ref{consecutivefinish}}).

\end{document}